\documentclass[12pt,preprint]{aastex}

\newcommand{\eref}[1]{equation (\ref{#1})}

\shorttitle{Peculiar Velocity and Deaberration}
\shortauthors{Menzies \& Mathews}

\begin{document}

\title{Peculiar Velocity and Deaberration of the Sky}

\author{D. Menzies and G. J. Mathews}
\affil{Department of Physics, University of Notre Dame, IN 46556, USA}
\email{dmenzies@nd.edu gmathews@nd.edu}

\begin{abstract}

Recent studies have found the earth's peculiar velocity to be significant in microwave background based tests for compact cosmic topology, and modifications to these tests have been proposed. Tests of non-gaussianity, weak lensing analysis and new tests using improved CMB data will also be sensitive to peculiar velocity. We propose here to simplify matters by showing how to construct a deaberrated CMB map to which any test requiring a Hubble flow viewpoint can be applied without further complication. In a similar manner deaberration can also be applied to object surveys used for example in topological searches and matter distribution analysis. In particular we have produced a revised list of objects with $z > 1.0$ using the NASA/IPAC Extragalactic Database.

\end{abstract}


\section{Introduction}

The peculiar velocity of the earth relative to the Hubble flow causes a direction aberration and a doppler shift of light that would be received by a coincident observer moving with the Hubble flow. Given the current precise estimate of the earth's peculiar velocity from CMB dipole measurement, \( (l,b) \approx (264.3^\circ, 48.0^\circ), \beta \approx 0.00123 \) \citep{lineweaver98CMBdipole}, the maximum angular error is \( \approx 0.07^\circ\) \citep{levin04aberratedCircles}. This is enough to cause concern for example in topological searches \citep{levin04aberratedCircles, calvao04aberratedCircles}, weak lensing and non-gaussianity analysis \citep{challinor01peculiarVelCMB} based on future CMB measurements. The impact of aberration on the CMB power spectrum is much less significant \citep{challinor01peculiarVelCMB}. Moreover, in analysis using object surveys, such as topological searches and matter distribution analysis, the aberration is already important since object positions are available to high accuracy. In the following, the term {\it aberration} shall include the doppler effect due to the peculiar motion of the earth relative to the Hubble flow, unless otherwise indicated. We shall summarize how to make these corrections, and present a new deaberrated table of objects with $z > 1$.

\section{CMB deaberration}

\begin{figure}[htbp]
\centerline{\resizebox{5in}{!}{\includegraphics{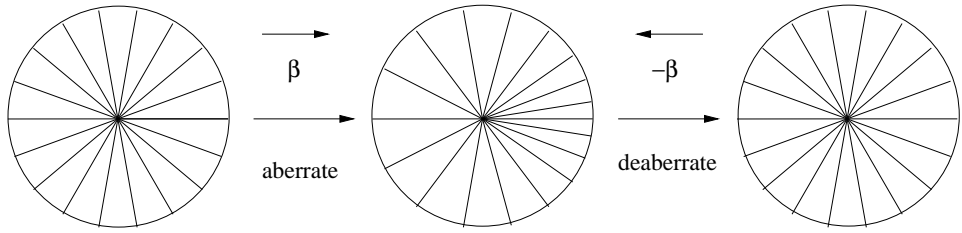}}}
\caption{\label{fig:picture} Deaberration restores light rays to unaberrated directions. }
\end{figure}

To deal with aberration, for example in the case of the {\it circle test} for evidence of cosmic topology \citep{cornish98circles}, modifications have been proposed to these tests \citep{levin04aberratedCircles, calvao04aberratedCircles}. An alternative approach, however, is to construct the view for an observer moving with the Hubble flow observer once, before applying any tests. Any further analysis can proceed as before unaltered, without the trouble of deriving and verifying modified tests that may require considerably increased computer time. This deaberration can be achieved by calculating the view of an observer moving with velocity \(-\vec{\beta}\) relative to the earth, where \(\vec{\beta}\) is the peculiar velocity of the earth relative to the Hubble flow. In other words, we aberrate our earth view with a velocity \(-\vec{\beta}\). This scheme is illustrated in figure~\ref{fig:picture}.

A light ray, viewed along the line of sight \(\hat{n}\) with a direction 3-vector \(-\hat{n}\), is aberrated by velocity \(\vec{\beta}\) as follows, \citep{calvao04aberratedCircles}:
\begin{equation}
\label{eq:vecs}
\hat{n}'= \frac{\hat{n}+[(\gamma-1)\hat{\beta}.\hat{n}+\gamma\beta]\hat{\beta}}
               {\gamma(1+\vec{\beta}.\hat{n})}
\quad , \quad \gamma = (1-\beta^2)^{-1}
\quad ,
\end{equation}
where the primed co-ordinate system refers to the aberrated (earth) viewpoint.
 A uniform black-body temperature field, \(T\), aberrates \citep{peebles68aberrate} into a blackbody temperature field \( T'(\hat{n}') \) according to
\begin{equation}
T'(\hat{n}')= \frac{T}{\gamma(1-\vec{\beta}.\hat{n}')}  \quad .
\end{equation}
To aberrate a general temperature field \(T(\hat{n})\) we substitute \(T\) with \(T(\hat{n})\), since light viewed along \(\hat{n}'\) in the aberrated view is viewed along \(\hat{n}\) in the unaberrated view, where \(\hat{n}, \hat{n}'\) are related by \eref{eq:vecs}.
\begin{equation}
\label{eq:T1}
T'(\hat{n}')= \frac{T(\hat{n})}{\gamma(1-\vec{\beta}.\hat{n}')}
\end{equation}
Consequently, by aberrating with velocity \(-\vec{\beta}\) the equation for the deaberrated temperature field is obtained:
\begin{equation}
\label{eq:T2}
T(\hat{n})= \frac{T'(\hat{n}')}{\gamma(1+\vec{\beta}.\hat{n})}
\quad ,
\end{equation}
where $\hat{n}'$ is a function of $\hat{n}$ given by \eref{eq:vecs}. Note that as a consistency check, \eref{eq:T1} can be substituted in the righthandside of  \eref{eq:T2} to give 
\( {T(\hat{n})}/[\gamma^2(1-\vec{\beta}.\hat{n}')(1+\vec{\beta}.\hat{n})] \),
which simplifies to \( T(\hat{n}) \) using \eref{eq:vecs}.

Given a pixelised map \eref{eq:T2} can be used to construct a deabearrated map provided a way is found to interpolate between pixels. It is reasonable to do this because of the spread beam profile of the observing instrument ensures the map is nearly bandlimited. The interpolation could be achieved using a spherical harmonic analysis of the map, $T'(\hat{n}') = \Sigma{ a'_{l m} Y_{l m}(\hat{n}')}$, calculated for instance using HEALPIX, \citep{healpix}. Such interpolation does not account for the incomplete sky coverage however. Localised interpolation could be achieved for example by using bicubic interpolation, but some uncertainty would remain.

It would be better to construct the deaberrated map directly from the raw time-ordered data produced by the observing instrument. Pixelization occurs at the end of this process and so the need to interpolate is eliminated. Typically CMB instruments produce output from several spectral channels, from which temperature is calculated, see the WMAP experiment for example \citep{wmapDataProcess}. The direction $\hat{n}'$ of a raw observation should be modified to its deaberrated direction $\hat{n}$. This is achieved by aberrating \(\hat{n}'\) with velocity \(-\vec{\beta}\) so that \eref{eq:vecs} transforms to  
\begin{equation}
\hat{n}= \frac{\hat{n}'+[(\gamma-1)\hat{\beta}.\hat{n}'-\gamma\beta]\hat{\beta}}
               {\gamma(1-\vec{\beta}.\hat{n}')}
\quad .
\label{eq:n}
\end{equation}
The deaberrated temperature, using \eref{eq:T1}, is
\begin{equation}
T(\hat{n}) = \gamma(1-\vec{\beta}.\hat{n}') T'(\hat{n}')
\end{equation}
The deaberrated data can then be used to construct a pixelized map in the same way a map would be constructed without deaberration.

\section{Deaberrating astrophysical objects}

In the case of objects we first wish to know the deaberrated line of sight object direction \(\hat{n}\) from the observed aberrated object direction \(\hat{n}'\). This has already been given in \eref{eq:n}.
The doppler effect when aberrating line of sight \(\hat{n}\) is \citep{calvao04aberratedCircles}
\begin{equation}
\label{eq:nu}
\nu' = \nu \gamma(1+\vec{\beta}.\hat{n})
\quad ,
\end{equation}
The deaberrated frequency, in terms of the observed \(\hat{n}'\), is then
\begin{equation}
\nu = \nu' \gamma(1-\vec{\beta}.\hat{n}')
\quad ,
\end{equation}
and the desired deaberrated spectral redshift is
\begin{equation}
z = \nu_0 / \nu - 1 = (\nu_0/\nu') / [ \gamma(1-\vec{\beta}.\hat{n}')] -1 = (z'+1) / [ \gamma(1-\vec{\beta}.\hat{n}')] -1
\quad ,
\end{equation}
where \(\nu_0\) is the unredshifted spectral line frequency.\\
The aberration of the photon number density spectrum from  \citep{peebles68aberrate}  is
\begin{equation}
n'(\nu',\hat{n}') = n(v,\hat{n}) (\nu' / \nu)^2
\quad .
\end{equation}
So the intensity spectrum \(I(\nu,\hat{n})=\hbar \nu n(v,\hat{n}) \) aberrates to
\begin{equation}
I'(\nu',\hat{n}') = I(v,\hat{n}) (\nu' / \nu)^3
\quad .
\end{equation}
Hence the deaberrated spectrums are
\begin{eqnarray}
n(\nu,\hat{n}) &=& n'(v',\hat{n}') / [\gamma(1+\vec{\beta}.\hat{n})]^2 \\
I(\nu,\hat{n}) &=& I'(v',\hat{n}') / [\gamma(1+\vec{\beta}.\hat{n})]^3
\quad ,
\end{eqnarray}
where $\nu'$ is given by \eref{eq:nu} and $\hat{n}'$ by \eref{eq:vecs}.

Using these formulae, we have prepared a table of deaberrated objects from an initial table of all objects with \(z > 1.0\) taken from NASA/IPAC Extragalactic Database \footnote{\url{http://nedwww.ipac.caltech.edu}}. Our deaberrated table can be found at \url{www.nd.edu/\(\sim\)dmenzies/deab}. For illustration, Table~\ref{tab1} summarizes the 50 objects with the largest angular aberration correction (but low redshift correction).

\section{Further peculiarities}

For greater accuracy we must account for the variation in peculiar velocity caused by the earth's rotation and movement about the sun, which is \(\approx 5\%\) of the total. To do this either the original observation must have been corrected for this variation, or we must know the time when the observation was taken and add to the average value of $\vec{\beta}$ the velocity of the earth relative to the sun at this time.

\section{Summary}

Formulae have been given for constructing deaberrated CMB maps and object surveys. It is suggested these could be applied one time prior to an application of tests that require a viewpoint that is at rest relative to the comoving Hubble flow. Otherwise each process must be modified and will suffer an increase in complexity and computational cost. Processes currently of interest are searches for non-trivial compact topology, and in the future tests of non-gaussianity, and analysis of weak lensing. Other tests will also undoubtedly come as the precision of cosmological data increases.

\acknowledgments

The authors wish to thank the referee for finding errors and suggesting improvements. Work at the University of Notre Dame supported by the U.S.
Department of Energy under research grant DE-FG02-95-ER40934, and a University of Notre Dame Center for Applied Mathematics (CAM) summer fellowship. This research has made use of the NASA/IPAC Extragalactic Database (NED) which is operated by the Jet Propulsion Laboratory, California Institute of Technology, under contract with the National Aeronautics and Space Administration.

\clearpage

\clearpage

\begin{deluxetable}{lrrrrr}
\tablewidth{0pt}
\tablecaption{Corrected galactic co-ordinates (in degrees) and redshift for the 50 most aberrated objects with $z > 1$. $\cos{\Delta\theta} = \hat{n} . \hat{n}'$\label{tab1}} 
\tablehead{
\colhead{Name}                   & \colhead{l}             &
\colhead{b}                        & \colhead{z}             &
\colhead{$\Delta \theta$}          & \colhead{$\Delta z$}  }
\startdata

\verb#ERO J164023+4644.0# & 72.67080 & 41.39394 & 1.049001 & 0.07047 & 0.000001\\
\verb#PKS 0230-790# & 297.14998 & -37.26604 & 1.069995 & 0.07047 & -0.000005\\
\verb#[MYF99] J0917+8142c# & 130.71970 & 31.78057 & 1.070000 & 0.07047 & 0.000000\\
\verb#PC 1639+4628# & 72.19289 & 41.26659 & 1.304997 & 0.07047 & -0.000003\\
\verb#[HB89] 1831-711# & 323.73096 & -24.51449 & 1.356005 & 0.07047 & 0.000005\\
\verb#[MYF99] J0923+8149a# & 130.46658 & 31.89165 & 1.389999 & 0.07047 & -0.000001\\
\verb#RX J1541.2+7126# & 106.64072 & 39.82372 & 1.418003 & 0.07047 & 0.000003\\
\verb#[HB89] 1700+180# & 38.16213 & 31.73087 & 1.423991 & 0.07047 & -0.000009\\
\verb#2MASSi J0453023-333359# & 235.69337 & -38.26683 & 1.800004 & 0.07047 & 0.000004\\
\verb#[MYF99] J0917+8142a# & 130.72920 & 31.79118 & 1.859999 & 0.07047 & -0.000001\\
\verb#[MYF99] J0917+8142b# & 130.72479 & 31.78714 & 1.869999 & 0.07047 & -0.000001\\
\verb#HE 0442-4445# & 249.88261 & -40.97861 & 1.918008 & 0.07047 & 0.000008\\
\verb#PMN J0459-2330# & 223.98060 & -34.50293 & 1.989999 & 0.07047 & -0.000001\\
\verb#ABELL 2219:[FBB2002] 1# & 72.69397 & 41.42281 & 4.068000 & 0.07047 & 0.000000\\
\verb#ABELL 2219:[FBB2002] 2# & 72.67086 & 41.41442 & 4.444999 & 0.07047 & -0.000001\\
\verb#ABELL 2219:[FBB2002] 3# & 72.62727 & 41.42965 & 4.654002 & 0.07047 & 0.000002\\
\verb#[HB89] 1642+409# & 64.92009 & 40.66151 & 1.300017 & 0.07047 & 0.000017\\
\verb#[HB89] 0448-392# & 242.74590 & -39.67631 & 1.302014 & 0.07047 & 0.000014\\
\verb#87GB 160847.5+654032# & 98.32452 & 40.82205 & 1.393985 & 0.07047 & -0.000015\\
\verb#[HB89] 1642+411# & 65.15790 & 40.69027 & 1.436017 & 0.07047 & 0.000017\\
\verb#PMN J0438-4728# & 253.64664 & -41.76199 & 1.444988 & 0.07047 & -0.000012\\
\verb#[HB89] 1643+406# & 64.45374 & 40.58646 & 1.451012 & 0.07047 & 0.000018\\
\verb#HS 0621+6738# & 147.11385 & 22.61153 & 1.588012 & 0.07047 & 0.000012\\
\verb#[HB89] 1642+412# & 65.22351 & 40.64394 & 1.970004 & 0.07047 & 0.000010\\
\verb#[HB89] 0447-395# & 243.11944 & -39.80981 & 1.980002 & 0.07047 & 0.000008\\
\verb#CTS 0514# & 254.19202 & -41.27335 & 2.369995 & 0.07047 & 0.000016\\
\verb#2MASSi J0446589-414601# & 246.06335 & -40.36583 & 2.700013 & 0.07047 & 0.000013\\
\verb#[HB89] 1640+471# & 73.13984 & 41.19657 & 2.763979 & 0.07047 & -0.000021\\
\verb#2MASSi J0445327-404848# & 244.79364 & -40.58809 & 3.269976 & 0.07047 & -0.000024\\
\verb#PC 1640+4628# & 72.17494 & 41.11833 & 3.694976 & 0.07047 & -0.000024\\
\verb#PC 1636+4635# & 72.37277 & 41.74207 & 1.2220051 & 0.07047 & 0.000012\\
\verb#[HB89] 1642+410 NED02# & 64.98530 & 40.79491 & 1.240010 & 0.07047 & 0.000016\\
\verb#[HB89] 1642+401# & 63.76762 & 40.56944 & 1.268022 & 0.07047 & 0.000022\\
\verb#[HB89] 1642+410 NED03# & 65.03083 & 40.77975 & 1.370022 & 0.07047 & 0.000022\\
\verb#[HB89] 1643+400# & 63.74611 & 40.48168 & 1.877022 & 0.07047 & 0.000022\\
\verb#CTS 0648# & 258.06942 & -41.39991 & 2.940036 & 0.07047 & 0.000036\\
\verb#PKS 0454-234# & 223.67308 & -34.96014 & 1.002978 & 0.07047 & -0.000022\\
\verb#3C 305.1# & 114.89062 & 38.26538 & 1.132022 & 0.07047 & 0.000022\\
\verb#CXOU J031015.9-765131# & 293.55532 & -37.68363 & 1.187022 & 0.07047 & 0.000022\\
\verb#HE 0435-5304# & 261.02111 & -41.44491 & 1.231026 & 0.07047 & 0.000026\\
\verb#[HB89] 1642+400# & 63.68701 & 40.64412 & 1.377027 & 0.07047 & 0.000027\\
\verb#CADIS 16h-0604# & 85.22938 & 42.44569 & 1.430025 & 0.07047 & 0.000025\\
\verb#[HB89] 1641+411# & 65.10552 & 40.85849 & 1.570027 & 0.07047 & 0.000027\\
\verb#[HB89] 1643+395# & 63.01211 & 40.46099 & 2.145031 & 0.07047 & 0.000031\\
\verb#[HB89] 1641+410# & 64.98338 & 40.86438 & 2.385037 & 0.07047 & 0.000037\\
\verb#MS 1006.3+8212# & 129.02387 & 33.12097 & 2.410037 & 0.07047 & 0.000033\\
\verb#CADIS 16h-0330# & 85.08463 & 42.44283 & 2.410033 & 0.07047 & 0.000033\\
\verb#HS 1649+3905# & 62.59049 & 39.32292 & 3.165949 & 0.07047 & -0.000051\\
\verb#CADIS 16h-0780# & 85.05864 & 42.47427 & 3.720048 & 0.07047 & 0.000048\\
\verb#CADIS 16h-2028# & 85.17524 & 42.54781 & 1.130027 & 0.07047 & 0.000027\\

\enddata
\end{deluxetable}

\end{document}